\documentclass[aps,twocolumn]{revtex4}

\usepackage{graphicx}
\usepackage{color}

\begin{document}

\bibliographystyle{apsrev}

\title{A Workshop that Works}

\author{Nicol\'as Yunes}

\affiliation{Department of Physics, 
  Montana State University, Bozeman, MT 59717}
  
\author{Joey Shapiro Key}

\affiliation{Center for Gravitational Wave Astronomy, Department of Physics and Astronomy, 
  University of Texas at Brownsville, Brownsville, TX 78520}

\date{\today}

\begin{abstract}
The main goal of a scientific workshop is to bring together experts in a specific field or related fields to collaborate, to discuss, and to creatively make progress in a particular area. The organizational aspects of such a meeting play a critical role in achieving these goals. We here present suggestions from scientists to scientists that hopefully help in organizing a successful scientific workshop that maximizes collaboration and creativity.
\end{abstract}

\pacs{}
\maketitle

\section{Introduction}
The interconnectivity of the modern world has allowed for almost instant access to publications and contact among scientists across the globe. We have come a long way since the early days of the internet with electronic bulletin boards. Today, Skype, Google Chat and many other telecommunication platforms allow for instant conversations with colleagues, as well as remote attendance at meetings and even conferences. For example, the presentation that announced the first possible detection of the Higgs particle was streamed live over the internet, so that people all over the world could get access to this information instantly.  

In spite of this interconnectivity that we take for granted, personal, face-to-face, scientific meetings have remained popular and one of the most productive and effective ways to develop collaborations. Typically, such meetings can be divided into \emph{conferences} and \emph{workshops}. The main goal of conferences is to bring together experts in a rather wide range of areas to disseminate new results. Because of this, conferences are usually much larger than workshops, involving hundreds or thousands of scientists, with a very large number of short talks (of duration ${\cal{O}}(10)$ minutes), a few long, plenary talks, and rather short breaks between talks.   

Workshops, on the other hand, usually have a very different goal: to bring together a smaller number of experts in a specific field or related fields to encourage collaboration, creativity, and progress during and following the workshop. Because of this, workshops are usually much smaller than conferences, perhaps involving less than one hundred scientists, with usually no short talks, a very small number of long talks, and ample time for discussion. 

The level to which a workshop achieves such goals is a sensitive function of its organization.  We here summarize a few suggestions on how to organize a successful workshop, distilled from our own experiences attending and organizing workshops, and discussions with other organizers and attendees. Scientific workshops come in a variety of forms and by no means do the suggestions that follow exhaust all possible ways to organize a workshop. In fact, as we will discuss below, there are many roads to success. The goal of this paper is to provide some guide to hopefully help workshop organizers (particularly young faculty and postdocs) when planning future meetings.  

The motivation for this paper comes from requests from colleagues, who have identified a need for such material. Although we searched the literature, we have not found a concise paper on this topic, written by \emph{scientists for scientists}. Of course, any workshop organizer could read on successful organizational techniques written by other professionals, think about how to translate these techniques to scientific meetings, and then compare and contrast these with techniques applied in the past in other meetings. This, however, requires careful research, planning, and preparation, which takes quite a bit of time and effort. 

The intended audience of this note is primarily young scientists that have been tasked with organizing a workshop for the first time. Such scientists may not have the time (or desire) to research and compile information on how to best organize a workshop. Experienced scientists, ie.~those who have organized workshops before, may find this note unnecessary. But of course, he who knows the answer to a problem usually thinks the answer is obvious. Hopefully, this note will present ideas that may not have occurred even to experienced scientists.

\section{Workshop Objectives}

In this section, we discuss how to define workshop objectives and then proceed to list and describe a set of tips that are useful when organizing a workshop. We will concentrate on \emph{workshops} only; there is not a one-to-one mapping between the tips presented here and those that are useful for conference organization. 

\subsection{Defining Objectives}
\label{subsec:objectives}

The first step in organizing a successful scientific workshop is to set \emph{workshop objectives}. This may sound obvious, but it is tremendously important because the rest of the planning follows directly from the stated objectives. In other words, the organizers must first determine (very early on) what will define success so that they can plan for it. The definition of success is not unique and it depends on the type of workshop one wishes to organize, the topic of the workshop, etc. 

In our interviews, we have found that some of the most common goals are to create a workshop that possesses
\begin{enumerate}
\item[(i)] an open and friendly atmosphere, 
\item[(ii)] ample opportunity for discussions, 
\item[(iii)] participants with high and distinct levels of expertise, 
\item[(iv)] incentives for participation.  
\end{enumerate}
All of these objectives are interconnected and support each other. Keeping these objectives in mind can help in the planning of other organizational details.
 
\subsection{Objective-Centered Organization}    

The best organizational techniques will follow from the organizers' objectives. Below, we list some organizational suggestions that support the objectives of Sec.~\ref{subsec:objectives}. 

\subsubsection{Topic Selection}

Careful consideration must be given to the selection of the workshop topic. The first issue to consider is that of \emph{broadness}. If one picks too broad of a topic, then the duration of the workshop may not be long enough. Long workshops with the same participants throughout are usually less successful because one runs the risks of ``burning the participants out''. Scientists have other administrative and teaching responsibilities at their own institutions that they must attend to. Usually, one can get away from these for some time, but unless one is on an extended leave or on sabbatical, usually one must return to these responsibilities after a short time. Moreover, long workshops run the risk of becoming repetitive, with participants either repeating each other or eventually losing motivation and interest, which usually results in useless discussions. On the other hand, too narrow of a topic may lead to a very small workshop that resembles more a large group meeting than a true workshop~\footnote{Sometimes, very narrow topics are desirable, as we will discuss below when describing \emph{busy days}}. When picking the breadth of a given topic, it is usually best to strive to strike a balance.

Another issue one must consider is that of \emph{timing} and \emph{community interest}. Having multiple workshops on a given particular topic organized a few months from each other is not desirable unless they are planned jointly and organized in a series. Similarly, picking topics that only a very small subset of the scientific community is interested in will lead to too small a workshop. Identifying the right topic of the right broadness, with the right timing and the right level of interest will strongly aid in attracting high-level participants, fulfilling objective (iii).  

A final critical organizational choice one must make is that of workshop \emph{duration}. One must strike a balance between a very short meeting (1 day) versus a meeting that is too long (4 days or longer). The sweet spot seems to be around $2$ or $3$ days. Workshops of this duration are long enough to incite discussion and allow for new ideas to emerge, while short enough to not exhaust participants and to fit in with busy schedules. The duration of the workshop, of course, is directly tied to the breadth of the workshop. Long-term workshops are also laudable enterprises, but they are inherently distinct from the type of meetings we are discussing here.

\subsubsection{Participant Selection}

The success or failure of the workshop depends strongly on the workshop participants. The participants should all be experts (or close to experts) in the topics of the workshop, but at the same time they should span a few different communities. On the one hand, a workshop with participants that are experts only on a very small subtopic might prevent original ideas and new approaches to problems to be discussed. On the other hand, a workshop with participants from widely different communities can prevent in-depth discussions on any particular area. Once more, a delicate balance must be struck between level of expertise and breadth of knowledge.  

Needless to say, all participants must maintain a cordial and collegiate, professional relation among each other in a workshop atmosphere. If participants cannot agree to these standards, then one runs the risk of creating a tense environment, that then suppresses discussion by all participants, affecting objectives (i) and (ii). The success of a workshop hinges on the participants feeling comfortable enough to discuss and share their thoughts. 

The organizers' ability to select participants can be enforced by inviting speakers first, and only after that inviting other participants directly. This can be done easily via email, once a (free) online registration page has been set up. Invitations should always be through \emph{personal} communications or emails and not through mass emails. Once a subset of people have been invited, the workshop can be advertised more broadly. One can draw from existing collaboration networks that may have been developed by the workshop organizers. Asking people to register for the meeting can also serve as a means to convince people to commit to the workshop.

Attracting good workshop participants can be achieved by providing certain incentives, which is particularly important if the workshop location is somewhat remote. In some order of priority, it is recommended that caffeinated beverages and perhaps some sort of food be provided during coffee breaks. Ideally, the organizers will also provide breakfast, to motivate participants to arrive to the workshop early and on time to listen to the first talk. If funding is available, organizers could also provide the incentive of covering hotel costs. This, however, can be very costly and, if limited funds are available, then organizers can choose to provide ``travel grants'' for the participants that need travel support the most. Such a selection can be made by asking for participants to complete a brief ``travel support'' application, if they need to, when they register online. 

Some workshop organizers have chosen to provide lunch during meetings, but we have found that this is usually not the best use of funds.  Lunch is a good time for a break and the effort needed to organize lunch for the whole group is usually better spent on other tasks. Long lunch breaks after an interesting morning session serve as excellent environments for smaller groups to continue interesting discussions and start collaborations. In general, we suggest to not charge for registration, unless funding is a severe problem. 
        
\subsubsection{Session Organization}       

The organization of the session structure is perhaps one of \emph{the} most important topics when organizing a workshop. There are many possible structures that can lead to a successful workshop. We will present below one particular structure that has led to successful workshops in the past. This structure requires that every workshop day be subdivided into two \emph{sessions} (a morning and an afternoon session), each with a few \emph{blocks} of a given duration, separated by extended coffee discussion breaks and lunch discussion breaks. 

The first thing to consider is when the morning session should start and when the afternoon session should end (ie.~the length of the workshop day). Given any set of participants, one is likely to find an admixture of ``morning people'' and ``not morning people.'' That is, workshops will contain a combination of people who are comfortable with very early morning sessions and those people who are not. It is thus important to find a balance that makes the mean happy. We have found that a good compromise is to have morning sessions that start around 9 am and afternoon sessions that end around 5 pm. Such ``late'' and ``early'' start and end times enhance participation and lead to more productive workshop hours.

The second consideration is the duration of the morning and afternoon sessions. This choice is controlled by the length of the lunch break. At the very least, the lunch break should last 1.5 hours. Shorter breaks usually lead to participants arriving late to the beginning of the afternoon session, which can be very disruptive. An ideal medium are breaks of about 2 hours, especially for workshop venues that lack lunch spaces near the workshop location. If one chooses the lunch break between 12:30 and 2:30 pm, this then automatically means the morning sessions go from 9:00am to 12:30am and the afternoon sessions from 2:30pm to 5:00pm. 

The third consideration is the subdivision of each session (morning and afternoon) into blocks. The number of blocks that fit into each session is determined by the desired length of each block and of the coffee discussion breaks. For reasons we explain below, blocks of about 1.5 hours appear to be ideal. Shorter blocks do not allow for in-depth discussions, while longer blocks lead to people leaving during the block for short breaks. Coffee breaks are not intended just for participants to have coffee or go to the bathroom, but also to continue in-depth discussions generated during the blocks. For this reason, coffee breaks of around 30 to 45 minutes are ideal.

Given these conditions, one arrives at the following possible breakdown: 
\begin{itemize}
\item Morning Session (9:00 am - 12:30 pm):
\begin{itemize}
	\item Block 1 (9:00 am - 10:30 am)
	\item Coffee Break (10:30 am - 11:00 am)
	\item Block 2 (11:00 am - 12:30 pm)
\end{itemize}
\item Lunch Break (12:30 pm - 2:30 pm)
\item Afternoon Session (2:30 pm - 5:00 pm):
\begin{itemize}
	\item Block 3 (2:30 pm - 4:00 pm)
	\item Coffee Break (4:00 pm - 5:00 pm)
\end{itemize}
\end{itemize}
Of course, as already argued, this is not the only possible breakdown, but rather one that works well. Notice that this structure was inferred by choosing certain criteria, directly associated with the objectives listed in Sec.~\ref{subsec:objectives}. Thus, each block's and break's duration is not chosen arbitrarily.   

At the beginning of every session (the morning and afternoon) it is helpful if the organizers remind the audience of the main objectives of that particular session. These objectives are to be developed and planned ahead of time, when deciding what each block is going to cover, so that the topics are well-connected. Such planning also aids in enhancing smooth transitions between blocks. One particularly good technique is to assign each block a set of questions on the particular topic that block is supposed to address. The organizers can then remind the participants of the topic of each block and the associated set of questions. 

A direct consequence of the breakdown above is a limit on the number of talks possible, which is a \emph{desired} objective, as ample time for discussion (objective (ii)) automatically implies fewer talks given a fixed workshop duration. One still has the freedom to choose how many talks to include in each block. We find that a single talk per block is truly all that a workshop should have. This is not because the talks should last 1.5 hours, but rather because one wishes each block to have lots of discussion embedded during each talk. Discussion is most naturally generated when participants have questions about material that is being presented. 

\subsubsection{Block Organization}       

A \emph{critical} ingredient in the organization of blocks under the paradigm described here is the use of ``participative talks''. These talks are those that are created with the goal to encourage discussion \emph{throughout} the talk, as opposed to after the talk in a separate discussion session. For such an interactive discussion situation to emerge naturally, the invited speaker must feel relaxed and willing to devote time during the talk to address questions and allow discussion. This, in turn, can only occur if the speaker has prepared \emph{few} slides (much, much fewer slides than for a 1.5 hour traditional talk). We have found that requesting 30-45 minute talks from the invited speakers is enough to fill up an entire 1.5 hour block that includes discussion. 

The organizers must stress and explain to the speakers the participative talk format. Nobody wishes to hear a sequence of 1.5 hour talks at a workshop. When this occurs, workshops begin to resemble conferences instead of collaborative and creative meetings. To prevent this, the organizers could attempt to ask to see the slides ahead of time, which then allows them to suggest cuts, if too many slides have been generated. This, however, can be very difficult to enforce as many speakers choose to prepare their presentations at the last minute. A perhaps better alternative is to clearly describe what a participative talk is supposed to be like and to remind speakers of this several times prior to the workshop.  

Another critical element in a participative talk is the level of complexity of the material presented. If the material is too technical, then one runs the risk of ostracizing the audience, thus preventing discussions. To ensure speakers prepare appropriate talks, the organizers could remind the presenters that these talks are not about pushing their personal research or agenda, or to impress the audience. The talks are meant to stimulate discussion, and thus, they should be clear and to the point, with a minimum use of technical jargon. One must avoid for only a small fraction of the participants to understand just a small fraction of the discussions that take place.

Participative talks are more useful in generating discussion than separate discussion sessions because of the way discussion usually emerges. Participants will  have questions and will want to discuss material while the speaker is presenting it. It is much easier to introduce these questions and discussion at that time than wait until a formal discussion session later in the day. Such questions then also allow other participants to understand the material better and to follow the discussions. Formal discussion sessions can easily degenerate into ``yet another talk'' given by the moderator or the discussion panel.  

Given the limited number of talks, each invited speaker should ideally cover a separate, non-overlapping topic that is carefully planned ahead of time. Organizers have the privilege and responsibility to select these topics carefully and in a coherent manner. Usually, speakers tend to recycle old talks, with perhaps a minimal modification that introduces new results. To avoid this, organizers can ask speakers to address new ideas or present questions, even at the beginning of the talk, rather than answers, about topics they think are worth studying. To avoid overlaps, it is useful for the organizers to put all speakers in contact with each other sufficiently ahead of the workshop. This way, speakers can discuss with each other what topics each of them will present to avoid overlaps. If speakers manage to finish their talks prior to the workshop, organizers may even make these talks available to all speakers to avoid overlaps.  

Each block should also have a carefully assigned moderator or ``chair'' to encourage discussion. If not enough discussion is being generated by the speaker's talk, then it is the job of the moderator to ``break the ice'' by asking questions until other participants join in. The moderator must also have the courage to stop discussions in the very rare situation when the latter have gone on for too long. For this reason, the moderator must be somebody familiar with the topic the speaker will discuss, while at the same time willing to ask questions in public. Just as speakers, moderators are not to be chosen at random or at the last minute.  

\subsubsection{Coffee Break Discussions}        

An important organizational element of workshop sessions is the time between blocks. This is precisely when participants mingle, draw on whiteboards, explain difficult concepts to each other, and discuss freely. There should be enough time in these coffee breaks to allow for such discussions and, in particular, to prevent these discussions from being interrupted abruptly by the end of the coffee session. Ideally, each coffee break session is at least 30-45 minutes long. 

Flexibility when starting and ending coffee breaks is also of utmost importance. Sometimes blocks finish a bit earlier than prescribed or coffee breaks seem to go for a bit longer. This is perfectly fine. The role of the organizer is to encourage discussion, collaboration, and creativity and never to stifle it. Ending or starting sessions prematurely at the cost of killing discussions should be avoided whenever possible.  

Discussions can be further enhanced if the beverages and food associated with the coffee breaks are served in the right environment. Ideally, coffee breaks would occur in the room adjacent to where the workshop is taking place. Using the same room can cause disruptions, as staff set up tables and prepare refreshments. If refreshments are served in the room adjacent to the workshop room preparations can start before sessions are over without disrupting discussions. 

Coffee break rooms should also be well-equipped to enhance collaboration. This means making available plenty of writing material (paper and pens), tables and chairs, as well as white boards. Many times white board discussions naturally emerge during coffee sessions, and this can be very productive. 
        
\subsubsection{Venue Selection}        

The venue for a workshop can greatly aid in ensuring its success. The impact of the venue is sometimes underestimated, but we have found (somewhat anecdotal) evidence, in fact, for the contrary. Workshops where the meeting rooms are vast, large, and imposing seem to lead to less discussion than medium size rooms, where all participants are physically close to each other. Moreover, large venues are prone to generate sub-discussions that occur simultaneously and in parallel to the main discussions of the session. This is counter-productive and it isolates participants instead of creating unity. 

Ideally, the meeting room where the workshop takes place is the same during all days of the workshop. This is facilitated by organizing workshops of intermediate duration (2-3 days), as opposed to multi-week endeavors. Of course, participants do not usually mind if they have to move from one room to another between workshop days, but sometimes this can create confusion if the room migration is non-trivial. Participants do mind if the venue is too far away from hotels. In this case, either a closer venue must be identified or a shuttle service should be provided to transport participants to the workshop. For all of these reason, the venue should be secured \emph{very early on} (e.g.~at least 6 months before the meeting) in the organization of a workshop. 

\subsubsection{Technical Infrastructure}        

Collaboration tools and equipment should be placed in the meeting room and be adequate for the size of the meeting. This means in particular securing sufficiently large tables for participants to place their laptops and notes on. Notepads could, for example, be provided freely as part of the registration package. If possible, tables should be arranged such as to encourage conversation. Tables in a $\Lambda$ pattern or great arcs can accomplish this, while allowing everybody to see the projection screens clearly. Moreover, meeting rooms should also have several well-illuminated whiteboards with bright markers. Participants and speakers often wish to add or explain material on whiteboards, making the latter of utmost importance. 

Projectors and pointers should be selected ahead of time and tested for brightness in well-illuminated rooms. University physics departments usually have projectors and pointers for colloquia that are much brighter than those that a venue for rent could provide. Usually, universities are willing to lend this equipment to faculty for free. Finally, power cords should be run across the workshop room for participants to charge their laptops, and high-speed wi-fi access should be provided. The latter should be easily accessible, preferably with an open network. Wi-fi access, unfortunately, tends sometimes to distract participants, which is why sometimes organizers would rather such access were not provided. However, the internet can also aid when searching for scientific material on the internet during discussion sessions, which can help in clarifying discussions. 

A workshop website should be constructed to serve as a hub to collect information. The website should contain a list of participants, schedule, information about accommodation and travel, as well as nearby restaurants, and possibly directions from nearby airports to hotels. Ideally, one would collect all presentations given at the workshop and upload them to the workshop website. If possible, one could also record the discussion sessions and upload these too. This last idea requires proper placement of relatively high-quality, environmental microphones to record the discussions and may not be available in the rented venue. Moreover, the organizers should ask the participants for permission prior to recording and uploading their presentations to the web. If this means the speakers will withhold information or shy away from discussion, then the meeting should not be recorded.  

\subsubsection{Plan B}        

It is important to be flexible when organizing a workshop since (almost certainly) not everything will go as planned. One of the most common failures is for invited speakers to cancel in the last minute. One must be prepared for such ``mini-disasters'' and plan accordingly. For this reason, it is always very useful to identify one or two people ahead of time and ask them to be possible back-ups, in case speakers cancel or cannot make it to the workshop for some reason (weather being the most common one). 

One can try to resolve this problem by allowing invited speakers to give talks via the internet, for example through Skype. In our experience, this solution is not as effective as it sounds. Skype is an ideal collaboration tool, but it is difficult for a speaker to give a talk with this technology, primarily because of not being able to gauge the audience's reactions in real time. Subconsciously speakers always adjust their talks in real-time in response to the audience. For example, if the audience looks confused, a speaker may choose to rephrase or explain a point further. Such real-time adjustments are impossible, or at the very least very difficult, through Skype. Moreover, tele-conferences of this type make it extremely difficult for participants to ask questions in real-time to interrupt and incite discussion. Without this very important element, participatory talks turn into traditional talks, which are much less productive for a workshop environment. 

Other disasters can of course also occur. Common problems include a last minute, forced change of venue, breakdown of collaboration equipment (like outlets, wi-fi, or projectors), absence of organizing members, or failure of the staffing members to provide adequate and timely refreshments during coffee breaks. All of these problems can be dealt with by the organizing committee easily, if solutions are thought of ahead of time, like the identification of a back-up venue.  
        
\subsubsection{Other Suggestions}

The collection of a group of visiting experts in a specific discipline is an excellent opportunity to reach out to the local community through an education or public outreach event.  Ideally, the organizers can recruit a colleague or collaborator with experience in outreach events to either advise on their planning or organize the event.  The goal is to have all of the participating scientists interact with the public or students on some level during an event that benefits both the audience and the scientists.  Public talks are common outreach events but many other formats can be successful, including school visits, panel discussions, and science caf\'{e}s.  Collaboration with colleagues in fields such as education, art, or history can lead to interesting and enjoyable outreach events.  It is a good experience for graduate and undergraduate students to participate in the planning and execution of outreach events as part of their own professional development.

A successful workshop can only occur if one attracts the appropriate audience. Of course, direct, personal invitations to special attendees are important, but advertising can also play a critical role. Mailing lists (such as professional society mailing lists) as well as community organized mailing lists (such as hyperspace for the relativity community) can be used to advertise meetings effectively. This advertisement should be written carefully and clearly including the date and location of the meeting, as well as a detailed description of what the workshop will be about. 

A conference dinner organized for all of the participants, whether provided as part of the workshop registration or organized as a no-host event, can often add to the social and collaborative nature of the workshop.  Informal dinner or evening events can often be a venue for the continued discussion of workshop or related topics as well as providing a different atmosphere for participants to get to know one another.  Organizers should consider the effort, cost, and accessibility of dinner or social events associated with the workshop. 

A good workshop attendance also hinges on when the meeting takes place. One should always try to avoid organizing a meeting in close proximity to other traditional meetings organized by the community of interest. For example, it makes sense to avoid organizing physics workshops on or around the same time as the March or April American Physical Society (APS) meetings. If the workshop is on topics related to astrophysics, one should make sure it does not overlap with the American Astronomical Society (AAS) meeting, with Aspen workshops, or with international general relativity or astrophysics meetings. Failing to do so will mean that potential speakers and workshop participants may choose not to attend the workshop and instead attend a more established and well-known conference. 

The particular season when the workshop takes place can play a major role in how well-attended it is. In the United States, the Fall (September to December) and the Spring semesters (January to May) are difficult since potential participants may have teaching duties. Winter can always be a problem due to weather delays, if the workshop is organized in a remote location. Summer can be an ideal time to organize meetings, but of course, this season is quite over-subscribed with other meetings and workshops.   

A successful workshop will attract a large number of scientists to the organizer's institution, and thus, the organizers may wish to offer visitors to stay after the workshop for some period of time. This serves two purposes. On the one hand, it allows the organizers to collaborate with the visitors more closely, since during the meeting itself, organizers are usually swamped with organizational duties. On the other hand, it allows graduate students and postdocs at the organizer's institution to establish and pursue new collaborations with the visitors beyond the duration of the meeting.   

Organizing a meeting is difficult and should be done \emph{early} enough and methodically. Usually, it is ideal to start organizing a small workshop about 8-12 months ahead of the meeting. After deciding on objectives and the format of the meeting, confirmations of attendance of the invited speakers should be secured, followed by confirmation of attendance of special attendees and the preparation of advertisement. One should simultaneously secure a venue early on to ensure accessibility to the best locations. Details that can be dealt with a few months prior to the meeting include arranging for registration packages (with notebooks, directions to restaurants from the workshop venue, wi-fi connectivity information, direction to hotels, etc) and refreshments.  

Graduate student help is also very important for a successful workshop organization. Participants that are new to the town where the workshop is being held sometimes need help to get around, find restaurants, or sometimes even the venue. Enlisting the help of graduate students can help with this, while introducing graduate students to well-known researchers and exposing them to the procedures for the successful organization of a workshop.         

Participatory workshops are successful if and only if participants participate. When workshops start, participation is minimal because people are sometimes shy or introverted. It is the responsibility of the block moderator and the organizers to break the ice and show everybody that it is ok to ask questions (many questions) and incite discussion. Participants may not be used to such a workshop format and may need to be shown that it is ok to interrupt and incite discussion. By breaking the ice, the organizers can help to ease the tension at the beginning of every meeting. A relaxed atmosphere will then naturally enhance the tendencies of participants to engage with the workshop. 

Proceedings and posters are usually not worth the effort and they take away from the informality needed in workshops to enhance participation. Organizers sometimes feel the need to have their participants write proceedings to have ``something to show for the meeting''. Historically, proceedings had equal weight to refereed scientific papers and served to let people all over the world know what new results had been presented. This is why proceedings were usually associated with conferences. Nowadays, however, proceedings have lost part of their utility given the advent of the internet and the arXiv. Many review papers currently exist that are up to date on several different topics, making proceedings usually redundant. The cost associated with writing such proceedings does not outweigh their benefit to the community. 

\subsubsection{Other Formats}
        
The above discussion has concentrated on a particular structure for a successful workshop. This structure was based on workshops we attended in the past, as well as workshops we organized, techniques used in business administration, and interviews conducted with workshop organizers and attendees. But by no means is this the only way to organize a successful workshop. One alternative format is that of a ``busy-meeting'' or ``hack-meeting''. These workshops are usually shorter than the ones described here and their goal is to solve a very specific problem with a small group of participants. Such hack-meetings can actually be embedded in larger workshops, in which case they turn into ``hack-days.'' When workshops are organized for large collaborations, for example, it is common for afternoon sessions to be hack-sessions, where multiple separate sub-groups of participants get together to correct a computer code, finish up a paper, or solve a theoretical physics problem. Such lightning sessions can be greatly successful. 

Other formats to each morning and afternoon session could also be used to enhance participation. A particular interesting alternative is the ``debate-format.'' In this exercise, two speakers are invited to debate on a particular topic, where one is asked to take one view, while the other must defend the opposing view (regardless of their own personal opinions on the topic). In science, it is common for conflicting results to arise in the literature. In such instances, it can be very illuminating to have the parties that discovered the conflicting results defend their position in a friendly by scientifically stimulating atmosphere. Of course, organizers must plan such sessions carefully, as sometimes the atmosphere can turn rather tense and be counter-productive. 

A particularly interesting variation is for all blocks to have a debate format. For this to succeed, of course, the organizers must be able to come up with enough questions or debate topics, as well as enough invited speakers who are capable of participating in such debates. This can be difficult if the workshop topic is not sufficiently broad, but it can work particularly well if one wishes to increase the breadth of the workshop.  

\section{Lessons Learnt}

The burden of whether a workshop is successful is ultimately on the organizers, not on the participants. Put another way, it is not the participants' fault if they are exhausted after a long day of talks and just not in the spirit to discuss topics any further. The organizers must consider all of these issues and realize that every organizational decision they make has a direct impact on the way the workshop turns out. It is thus the organizers' responsibility to plan ahead and provide participants with the tools to make the workshop successful. Because in the end, it is the participants' involvement that can either make or break a meeting. 

In order to quantitatively assess the success of a meeting, organizers should plan to collect data useful in determining the quality of the workshop. This can be achieved with \emph{anonymous} ``exit-surveys'', that can be completed at the end of the workshop. Simple questions like: ``Did you enjoy the meeting?'' or ``Would you attend another meeting of this type?'' can provide useful data to quantitatively study workshop success, for example, as a function of workshop size and venue selection. Ideally, data would be collected for many workshops with the goal of aggregating enough data to quantify the workshop parameters that maximize workshop success. 

\acknowledgments We would like to thank several graduate students, including Laura Sampson, Paul Baker, Katerina Chatziioannou, Dimitry Ayzenberg, Nicholas Loutrel, and Meg Millhouse, as well as postdoctoral researchers, including Kent Yagi and Antoine Klein, for helping us organize the workshop ``Gravitational Wave Tests of Alternative Theories of Gravity in the Advanced Detector Era'', held at Montana State University in Bozeman, Montana in April 2013. We would also like to thank several physicists and astrophysicists for discussions on this topic, including Pau Amaro-Seoane, Emanuele Berti, Vitor Cardoso, Neil Cornish, Pedro Ferreira, Jon Gair, Tyson Littenberg, Ed Porter, Leo Stein, and Carlos Sopuerta. NY acknowledges support from NSF grant PHY-1114374, PHY-1234826, PHY-1250636, as well as support provided by NASA grant NNX11AI49G, under sub-award 00001944. 


\end{document}